\magnification=1200
\hsize 14.9truecm \hoffset 1.2truecm
\vsize 22.2truecm\voffset .2truecm
\font \titolo=cmbx12 scaled\magstep0
\font \ninerm=cmr10 at 10truept
 at 10truept

\baselineskip=18truept 

\def \om {\omega}
\def \n {\eta}

\def \La {\Lambda}

\def \a {\alpha}

\def \ga {\gamma}
\def \sg {\sigma}
\def \da {\delta}
\def \ep {\epsilon}
\def \part {\partial}

\def \um {{1\over 2}}

\def\IR{{\hbox{{\rm I}\kern-.2em\hbox{\rm R}}}}
\def\IT{{\hbox{{\rm I}\kern-.2em\hbox{\rm T}}}}
\def\I{{\hbox{{\rm I}\kern-.2em\hbox{\rm I}}}}
\def\IH{{\hbox{{\rm I}\kern-.2em\hbox{\rm H}}}}
\def\IC{{\hbox{\kern-.2em\hbox{\bf C}}}}
\def\IZ{{\hbox{{\rm Z}\kern-.4em\hbox{\rm Z}}}}
\rightline{DFTT 37/97}
\rightline{YCTP-P10-97}
\rightline{ESI Preprint 475}
\rightline{gr-qc/9707032}
\vskip 3ex
\centerline{\titolo Constants of motion and the conformal 
anti - de Sitter  }
\centerline{\titolo algebra in (2+1)-Dimensional Gravity}

\vskip 0.7truecm

\centerline{V.~Moncrief%
\footnote{\raise2pt \hbox{\ninerm*}}{\ninerm Electronic address: 
~moncrief@hepmail.physics.yale.edu}}
\centerline{\it Department of Physics and Department of Mathematics}
\centerline{\it Yale University, 217 Prospect Street}
\centerline{\it New Haven, Conn. 06511, USA}
\vskip 0.7truecm
\centerline{J.E.~Nelson%
\footnote{\raise2pt \hbox{\ninerm**}}{\ninerm Electronic address: 
~nelson@to.infn.it}}
\centerline{\it Dipartimento di Fisica Teorica dell'Universit\`a di  
Torino}
\centerline{\it via Pietro Giuria 1, 10125 Torino, Italy}
\vskip 1.0 truecm

Constants of motion are calculated for 2+1 dimensional gravity with
topology $\IR \times T^2$ and negative cosmological constant. Certain 
linear combinations of them satisfy the 
anti - de Sitter algebra $\hbox{so}(2,2)$ in either ADM or holonomy 
variables. Quantisation is straightforward in terms of the holonomy 
parameters. On inclusion of the Hamiltonian three new global 
constants are derived and the 
quantum algebra extends to that of the conformal algebra 
$\hbox{so}(2,3)$. The modular group appears as a discrete 
subgroup of the conformal group. Its quantum action is generated 
by these conserved quantities.
\vskip 1ex
\leftline {\ninerm PACS number(s) \ 04.60.Kz, 04.60.Ds}
\vfill\eject
\centerline{\bf I. INTRODUCTION}
\vskip 0.7truecm

The Einstein equations for pure 2+1 gravity (with or without a 
cosmological constant), formulated for spacetimes having a compact 
Cauchy surface, can be reduced (\`a la ADM) to a finite dimensional, 
time dependent Hamiltonian system defined on the cotangent bundle of the 
Teichm\"uller space of the chosen (Cauchy) surface [1]. This ADM-reduced 
system admits sufficiently many independent conserved quantities (traces 
of holonomies) that one can, in principle at least, determine the 
evolution of the Teichm\"uller parameters and their conjugate momenta 
(i.e., the reduced ADM variables) by simply evaluating the conserved 
quantities in terms of them and the chosen ADM time coordinate, setting 
the resultant expressions equal to certain constants fixed by the 
desired initial conditions, and solving algebraically for the ADM 
variables as functions of the chosen constants and time. In other words, 
merely evaluating the conserved quantities in terms of the ADM phase 
space variables and time gives the solution of Hamilton's equations 
implicitly. This procedure can be carried out explicitly for the case of 
Cauchy surfaces diffeomorphic to the torus (the spherical case being 
essentially trivial and the higher genus case nearly intractable in 
practice).

The reduced ADM dynamics for the torus case can be quantised 
in a straightforward way by 
converting the Hamiltonian to a positive self-adjoint operator (the 
square root of a Laplace-Beltrami operator) acting on square-integrable 
functions on Teichm\"uller space (or, more precisely, on moduli space). 
The associated quantum dynamics has been studied in some detail by Puzio 
for the vacuum case [2]. One can ask, however, whether there is another 
approach to solving the quantum dynamics modelled on the classical 
technique outlined above. Could one define quantum analogues of the 
classical conserved quantities, set these equal to certain constant 
operators having appropriate commutation relations and solve the 
resulting formulae for the ADM operators (representing the quantised 
Teichm\"uller variables and their momenta) in terms of a set of fixed 
operators and time? This would amount to solving the Heisenberg 
equations of motion by a technique which closely parallels the solution 
of the classical Hamilton equations descibed above. Some guidance for 
doing this is provided by the classical Poisson bracket algebra of the 
conserved quantities which, presumably, one wishes to implement quantum 
mechanically if possible. In this paper we study this algebra and its 
possible quantum implementations from several points of view which are 
not necessarily equivalent at the quantum level.

This quantum mechanical implementation of the classical algebra is 
indeed successful provided we first express the conserved quantities in 
terms of certain holonomy parameters which were first introduced in [3]. 
The conserved quantities then take the form of purely quadratic 
expressions for which quantisation is straightforward. Having done this 
we then proceed to study the quantum action of the modular group upon 
the chosen representation of the conserved quantities. We also present a 
time dependent transformation relating the quantised holonomy parameters 
and the moduli and their momenta (i.e. the ADM variables) that was first 
discovered, for $\La = 0$, in [4] and extended to $\La < 0$ in [5]. 
However we should mention that the quantum dynamics 
may not be unitarily equivalent to that involving the (square 
root) Hamiltonian studied by Puzio [2]. At least we do not know of 
such a 
unitary correspondence and suspect that it may not exist since otherwise 
the simple formula given in [5] would encode within it the details 
of the spectrum of the Laplacian on moduli space. Thus, even though we 
present a solution of the Heisenberg equations of motion in terms of 
certain constant operators which are simply related to the quantum 
analogues of the conserved quantities discussed above, we do not claim 
that this is unitarily related to the Schr\"odinger problem posed by 
Puzio [2] whose solution therefore remains open.

The ADM approach to quantisation has the advantage of focusing on 
dynamics and it has a relatively clear correspondence to the classical 
theory. Its main disadvantages are that it requires the {\it a priori} 
choice of a time gauge (presumably breaking "general covariance" at the 
quantum level) and that it leads to nearly intractable dynamics (even 
classically) for the higher genus cases. An alternative approach to 
quantisation developed by Nelson and Regge [3,6-7] has the great 
advantages 
that it does not require the {\it a priori} choice of a gauge and that 
it remains tractable for the higher genus problems. Its principal 
disadvantage is perhaps that it does not (as yet) deal with dynamical 
questions and thus is more difficult to connect (in the correspondence 
limit) to the classical picture of spacetime as "space evolving in 
time". We hope that our present approach, however incomplete, may 
eventually lead to a reconciliation of these seemingly disparate 
approaches to quantum gravity in 2+1 dimensions. 

The structure of the paper is as follows. In Section II we derive a set 
of constants of the motion and express them in terms of both the ADM 
variables and the global holonomy parameters. In Section III we show 
that certain combinations of these constants satisfy the classical
anti - de Sitter Poisson bracket algebra. In Section IV we discuss 
the quantisation of this algebra and 
show how it is straightforward in terms of the holonomy parameters. In 
Section V the Hamiltonian is included in the algebra leading to 
three new global, quantum, constants of the motion. This extended 
algebra is isomorphic to the algebra of the conformal group 
$\hbox{SO}(2,3)$. In Section VI the action of the modular group 
is calculated and shown to 
be generated by precisely the (quantum) constants of the motion which 
generate a discrete subgroup of the conformal group. Our 
results are summarised in Section VII.

\vskip 0.7truecm

\centerline{\bf II. CONSTANTS OF THE MOTION}
\vskip 0.7truecm

As discussed in  [1], one can show that the vacuum spacetimes on  
$\IR \times T^2$ having negative cosmological constant $\Lambda$ and  
admitting a Cauchy surface of constant mean curvature $\tau$ are all  
spatially homogeneous.  In suitable coordinates their metrics can be  
written in the form

$$
ds^2 = - N(t)^2(dt)^2 + e^{2\mu (t)}(dx^1)^2 + e^{2\nu (t)} (dx^2 +  
\beta (t) dx^1)^2 \eqno(2.1)
$$

\noindent
where $x^1$ and $x^2$ are each periodic (with period = 1 for  
convenience) on $S^1$.

As in  [8] we introduce the orthonormal frame

$$
\eqalign{
e^{(0)} &= N(t)dt, ~~e^{(1)} = e^{\mu (t)} dx^1, \cr
e^{(2)} &= e^{\nu (t)} (dx^2 + \beta (t) dx^1) \cr
}  \eqno(2.2)
$$

\noindent
and compute the connection one-forms $\omega_{(a)(b)} =  
\omega_{(a)(b)\mu} dx^{\mu}$

$$
\eqalign{
\omega_{(1)(0)} &= - \omega_{(0)(1)} = A(t) dx^1 + C(t) dx^2, \cr
\omega_{(2)(0)} &= - \omega_{(0)(2)} = B(t) dx^1 + D(t) dx^2, \cr
\omega_{(1)(2)} &= - \omega_{(2)(1)} = {1 \over 2} e^{\nu (t) - \mu  
(t)} \beta_{,t} ~dt , \cr
}  \eqno(2.3)
$$

\noindent
where

$$
\eqalign{
A(t) &= {1 \over N(t)} (e^{\mu} \mu_{,t} + {1 \over 2} ~e^{2\nu -  
\mu} \beta \beta_{,t}) \cr
B(t) &= {1 \over N(t)} (\beta e^{\nu} \nu_{,t} + {1 \over 2} ~e^{\nu}  
\beta_{,t}) \cr
C(t) &= {1 \over N(t)} ({1 \over 2} ~e^{2\nu - \mu} \beta_{,t}) \cr
D(t) &= {1 \over N(t)} e^{\nu} \nu_{,t} .\cr
}  \eqno(2.4)
$$

\noindent
In terms of these quantities we introduce the duals

$$
\omega^{(a)} = {1 \over 2} ~\epsilon^{abc} \omega_{(b)(c)}  \eqno(2.5)
$$

\noindent
and a pair of "shifted connections" ~$\lambda^{\pm}{}^{(a)}$ ~defined  
by

$$
\lambda^{\pm}{}^{(a)} = \omega^{(a)} \pm \sqrt{- \Lambda} ~e^{(a)} .   
\eqno(2.6)
$$

\noindent
From  [9] we know that the traces of the $\hbox{SO}(1,2)$  holonomies 
of $\lambda^+$ and $\lambda^-$, defined for arbitrary closed loops in  
these spacetimes, are absolutely conserved quantities (i.e., they are  
gauge invariant and invariant under non-singular deformations of the  
loops within the vacuum spacetimes).

As in  [8] we compute this pair of traces for 3-different classes  
of loops represented respectively by the "a-loops" having $x^2 =  
constant$, the "b-loops" having $x^1 = constant$ and the "twisting  
loops" having $x^1 = {p \over q} ~x^2$.   The twisting loops do not  
yield new independent conserved quantities but rather give certain  
functions of those coming from the a- and b-loops.

Using a convenient representation of $\hbox{SO}(1,2)$ as in  [8] and 
[9] one extracts from the traces the following conserved quantities

$$
\eqalign{
C_1^{\pm} &=  (B \pm \sqrt{- \Lambda} e^{\mu})^2 + (-A \pm \sqrt{-  
\Lambda} ~e^{\nu} \beta)^2 , \cr
C_2^{\pm} &=  D^2 + (- C \pm \sqrt{- \Lambda} e^{\nu})^2 , \cr
}  \eqno(2.7)
$$

\noindent
(which come from the a- and b-loops respectively) and

$$
C_3^{\pm} =  (- A \pm \sqrt{- \Lambda} e^{\nu} \beta) (-C \pm \sqrt{-  
\Lambda} ~e^{\nu}) + D (B \pm \sqrt{- \Lambda} e^{\mu})  \eqno(2.8)
$$

\noindent
(which come from the twisting loops).

For the sake of easy comparison with earlier work on the $\Lambda =  
0$ problem we decompose the above expressions into the following  
equivalent set of conserved quantities

$$
\eqalign{
C_1 &= {(C_1^+ + C_1^-) \over 2} = A^2 + B^2 - \Lambda (e^{2\mu} + e^{2\nu} \beta^2) \cr
C_2 &= {(C_2^+ + C_2^-) \over 2} = C^2 + D^2 - \Lambda e^{2\nu} \cr
C_3 &= {(C_3^+ + C_3^-) \over 2} = AC + BD - \Lambda e^{2\nu} \beta \cr
C_4 &= {(C_1^+ - C_1^-) \over {4 \sqrt{- \Lambda}}} = B e^{\mu} - A e^{\nu} \beta \cr
C_5 &= - {(C_2^+ - C_2^-) \over {4 \sqrt{- \Lambda}}} = e^{\nu} C \cr
C_6 &= {(C_3^+ - C_3^-) \over {2 \sqrt{- \Lambda}}} = e^{\mu} D - e^{\nu} ~\beta ~C - e^{\nu} A \cr
}  \eqno(2.9)
$$

\noindent
In the limit that $\Lambda \rightarrow 0$ these quantities reduce to  
those defined in  [8].
\vskip 0.7truecm

\centerline{\bf A. ADM variables}
\vskip 0.7truecm

As in [8] we define new canonical coordinates $\left\{ q^i \right\}$  
by

$$
q^1 = \nu - \mu , q^2 = \beta , q^3 = \nu + \mu  \eqno(2.10)
$$

\noindent
and introduce their conjugate momenta $\left\{ p_i \right\}$ such  
that

$$
\sum_{i}^{} p_i q_{,t}^i = \sum \pi^{ab} g_{ab,t}  \eqno(2.11)
$$

\noindent
where $\left\{ g_{ab} , \pi^{ab} \right\}$ are the usual Arnowitt,  
Deser and Misner (ADM) canonical variables for the metric (2.1).  With  
these definitions $q^1$ and $q^2$ parametrize the conformal metric

$$
h_{ab} = {g_{ab} \over \sqrt{^{(2)}g}}  \eqno(2.12)
$$

\noindent
while $q^3$ parametrizes the spatial volume element, $e^{q^3} 
= \sqrt{^{(2)}g}$ and the mean  curvature $\tau$ is given by

$$
\tau = {p_3 \over e^{q^3}}  \eqno(2.13)
$$

By virture of spatial homogeneity the momentum constraints are  
satisfied identically while the Hamiltonian constraint now takes the  
form

$$
{\cal H} = {1 \over e^{q^3}} ({1 \over 2} p_1^2 + {1 \over 2} ~e^{-  
2q^1} p_2^2 - {1 \over 2} p_3^2) + 2 \Lambda e^{q^3}  .  \eqno(2.14)
$$

\noindent
The ADM super Hamiltonian is thus

$$
H_{super} = \int_{T^2}^{}{} ~N {\cal H} d^2x = N {\cal H}  \eqno(2.15)
$$

\noindent
from which one derives the relations

$$
\eqalign{
\dot{q}^1 &= {N \over e^{q^3}} p_1 ,  ~\dot{q}^2 = {N \over e^{q^3}}  
~e^{-2q^1} p_2 , \cr
\dot{q}^3 &= -  {N \over e^{q^3}} p_3  . \cr
}  \eqno(2.16)
$$

\noindent
Using (2.4) and (2.10-11) we can easily express the conserved quantities $C_1  
- C_6$ in terms of the canonical variables $\left\{ q^i, p_i  
\right\}$. First however we wish to ADM-reduce the dynamics by  
choosing the York time coordinate condition

$$
t = \tau = {p_3 \over e^{q^3}}  \eqno(2.17)
$$

\noindent
and by solving the constraint ${\cal H} = 0$ (2.14) for $e^{q^3}$ which  
serves to eliminate the pair $\left\{ q^3, p_3 \right\}$ from the  
dynamical equations.  With this choice of gauge the reduced, ADM,  
Hamiltonian is just the spatial volume
$$
H_{ADM} = \int_{T^2} d^2x \sqrt{^{(2)}g} = e^{q^3} = {1 \over  
\sqrt{\tau^2 - 4\Lambda}} ~\bar{H}  \eqno(2.18)
$$
where
$$
\bar{H} = \sqrt{p_1^2 + e^{-2q^1}p_2^2}  \eqno(2.19)
$$
is explicitly time independent.

Thus, as expected since volume is not conserved, the ADM Hamiltonian   
(2.18) is explicitly time dependent.  Here however the time dependence  
resides in a multiplicative factor which we can eliminate by simply  
changing the time variable from York time $\tau$ to a new time  
$t^{\prime}$ defined by

$$
\tau = \sqrt{-4 \Lambda} ~{\rm sinh}~ t^{\prime}  \eqno(2.20)
$$

\noindent
With this choice the reduced Hamilton equations are

$$
{dq^i \over dt^{\prime}} = {\partial \bar{H}  \over \partial p_i}  
~,~~ {dp_i \over dt^{\prime}} = - {\partial \bar{H} \over \partial  
q^i}  \eqno(2.21)
$$

\noindent
with $\bar{H}$ given by (2.19).

In terms of the foregoing definitions, the conserved quantities $C_1  
- C_6$ now take the form

$$
\eqalign{
C_1 &= {1 \over 2} ~e^{-q^1} \tau \left\{ (\sqrt{1 - {4\Lambda \over  
\tau^2}} \bar{H} - p_1)(1 + (q^2)^2 e^{2q^1}) - 2(q^2 p_2 - p_1)  
\right\} , \cr
C_2 &= {1 \over 2} ~e^{q^1} \tau\left\{\sqrt{1 - {4\Lambda \over  
\tau^2}} \bar{H} - p_1 \right\} , \cr
C_3 &= {1 \over 2} ~e^{q^1} \tau\left\{ q^2 (\sqrt{1 - {4\Lambda  
\over \tau^2}} \bar{H} - p_1) - p_2  ~e^{-2q^1} \right\} , \cr
C_4 &= {1 \over 2} \left\{ p_2 ~e^{- 2q^1} + 2 q^2 p_1 - p_2 (q^2)^2  
\right\} , \cr
C_5 &= {1 \over 2} ~p_2 \cr
C_6 &= p_1 - q^2 p_2 \cr
}  \eqno(2.22)
$$

\noindent
and reduce to those found previously [8] when $\Lambda \rightarrow  
0$.

While $C_1 - C_6$ are conserved quantities by construction one could  
verify this fact directly using the reduced Hamilton equations

$$
{dq^i \over dt^{\prime}} = {\partial \bar{H}  \over \partial p_i}  
~,~~ {dp_i \over dt^{\prime}} = - {\partial \bar{H} \over \partial  
q^i} , ~~i = 1, 2 .  \eqno(2.23)
$$

\noindent
We shall verify this in a different way below but mention here that  
one can simply use the constancy of the $C_i$'s to solve Hamilton's  
equations by purely algebraic means.  One can simply choose four of  
the independent $C_i$'s and solve them for the four canonical  
variables $\left\{ q^i, p_i \right\}$ in terms of $\tau = \sqrt{-  
4\Lambda} ~{\rm sinh} ~t^{\prime}$ and the four independent  
constants.  That this produces the solution to Hamilton's equations  
is equivalent to the fact that four of the $C_i$'s are functionally  
independent constants of the motion.
\vskip 0.7truecm

\centerline{\bf B. Holonomy parameters}
\vskip 0.7truecm

That the  $C_1  - C_6$ are time independent can be seen alternatively 
by reexpressing these quantities in terms of the global, time 
independent parameters $r_{1,2}^{\pm}$ of the traces of the 
$\hbox{SL(2,$\IR$)}$ holonomies  (Wilson loops)
[3,5], expressed as
$$\eqalign{R_1^{\pm}&= \cosh{r_1^{\pm} \over 2}\cr
R_2^{\pm}&= \cosh{r_2^{\pm} \over 2}\cr}\eqno(2.24)$$
where the subscripts 1 and 2 in (2.24) refer to two intersecting paths 
$\ga_1, \ga_2$ on $T^2$ ( the "a-loops" and "b-loops" of Section IIA, 
respectively) with intersection number $+1$. (A third  
holonomy, 
$R^\pm_{12} = \cosh{(r_1^{\pm} + r_2^{\pm})/2}$, corresponds to the 
path $\ga_1\cdot\ga_2$, the "twisting loops", which has intersection 
number $-1$ with  $\ga_1$ and $+1$ with $\ga_2$.)
In (2.24) the $\pm$ refer to the two copies of $\hbox{SL(2,$\IR$)}$
which appear in the decomposition of the spinor group of 
$\hbox{SO}(2,2)$ 
as a tensor product $\hbox{SL(2,$\IR$)}\otimes\hbox{SL(2,$\IR$)}$.

The holonomies (2.24) are the normalised traces of the 
hyperbolic-hyperbolic representation of $\hbox{SL(2,$\IR$)}$
(here hyperbolic means that the 
normalised trace is $>1$). This hyperbolic-hyperbolic
representation is necessary for the toroidal slices to be spacelike [10].
They completely solve all constraints in the alternative first order, 
formalism, and can be calculated directly from the classical solutions 
(2.2-3), or from the "shifted connections" (2.6) in the time 
gauge $N=1, N^i=0$. This 
gauge is equivalent, for the topology $\IR \times T^2$, to the York 
gauge of Section IIA. Explicitly,
$$
(r_{1,2}^{\pm})^2= \Delta_{1,2}^{\pm}{}^a~\Delta_{1,2}^{\pm}{}^b~\eta_{ab},
\quad  \eta_{ab} = diag(-1,1,1)$$
with $$
\Delta_{1,2}^{\pm}{}^a = \int_{\ga_1, \ga_2} \lambda^{\pm}{}^{(a)}
\eqno(2.25)$$
and $\lambda^{\pm}{}^{(a)}$ is given by (2.6).
Moreover, the holonomies (2.24) satisfy the nonlinear
classical Poisson bracket algebra [4,6]
$$\{R_1^{\pm},R_2^{\pm}\}=\mp{1\over 4\a}(R_{12}^{\pm}-
 R_1^{\pm}R_2^{\pm}) $$
$$\{R_1^{+},R_2^{-}\}= 0,\eqno(2.26)$$
where
$$ \a = {1 \over \sqrt{-  \Lambda}} > 0.$$

The algebra (2.26) is obtained by integration of the Poisson brackets
$$\{e^{(a)}_i({\bf x}),\om_{j (b)(c)}({\bf y})\}=- \um 
\ep_{ij}{\ep^a}_{bc} \da^2({\bf x-y}),\quad i,j= 1,2, 
\ep_{12}=1 \eqno(2.27) $$
along $\ga_1$ and $\ga_2$ [3]. When the traces $R_{1,2}^{\pm}$ are 
represented as in (2.24) it follows that the parameters 
$r_{1,2}^{\pm}$ satisfy the classical Poisson brackets
$$\{r_1^\pm,r_2^\pm\}=\mp {1\over\a}, \qquad \{r^+,r^-\}= 0 
\eqno(2.28)$$ 

The four real parameters $r_1^{\pm}, r_2^{\pm}$ appearing in (2.24)
are arbitrary, but in [5] it was shown that they are related, 
through a time-dependent canonical transformation, to the 
components of the moduli $m = m_1 +im_2$ and their momenta 
$\pi=\pi^1 +i \pi^2$ as follows%
\footnote{\raise2pt \hbox{\ninerm*}}{\ninerm In [5] $p= p^1+ip^2$ was 
used to denote the complex moduli momenta $\pi$. Here $p_1$ and $p_2$ 
denote the ADM momenta of Section IIA.}.

$$m= \left(r_1^-e^{it/\a} + r_1^+e^{-{it/\a}}\right) 
 \left(r_2^-e^{it/\a} + r_2^+e^{-{it/\a}}\right)^{\lower2pt%
 \hbox{$\scriptstyle -1$}} \eqno(2.29)$$

$$\pi= -{i\a\over 2\sin{2t\over \a}}\left(r_2^+e^{it/\a} 
 + r_2^-e^{-{it/\a}}\right)^{\lower2pt%
 \hbox{$\scriptstyle 2$}} \eqno(2.30)$$
where $m$ and $\pi$ are related to the ADM variables
$ q^1, q^2, p_1, p_2$ of Section IIA by 

$$m_1 = q^2, m_2 = e ^{-q^1}, \pi^1 = p_2, \pi^2 = -p_1 
e^{q^1}\eqno(2.31)$$

The parameter $t$ appearing in (2.29-30) is related to the extrinsic 
curvature $\tau$ by
$$\tau= - {\dot q}^3 
= -{2 \over {\a}}\cot {2t \over  {\a}} \eqno(2.32) $$ 
with $\tau$ monotonic in the range $t~ \epsilon~ (0, {\pi \a \over 2})$.

Since the $r_1^{\pm}, r_2^{\pm}$ are arbitrary  the moduli and momenta  
can have arbitrary initial data $m(t_0), p(t_0)$ at some initial time 
$t_0$. 

In [5] it was shown that the moduli $m_1, m_2$ with $m_2 > 0$ lie on a 
circle  
$$ (m_1 - c)^2 + m_2^2 = \vert m - c \vert^2 = R^2.\eqno(2.33)$$
It follows that the centre of this circle given by $m_2 =0, m_1 =c$. and 
its radius $R$ can be expressed 
in terms of the constants of Section IIA by differentiating (2.33) with 
respect to $t$ and using (2.16), (2.22) and (2.31)
$$c= -{C_6 \over 2C_5}, \qquad R^2 ={C_6^2 +4C_4C_5 \over 
4C_5^2}.\eqno(2.34)$$

The Poisson brackets (2.28) of the parameters $r_{1,2}^{\pm}$ can be 
used to 
calculate those of the moduli and their momenta. From  (2.29-30) we find
$$\{\bar m,\pi\}=\{m,\bar \pi\}=-2, \qquad \{m,\pi\}=
\{\bar m,\bar \pi\}=0\eqno(2.35)$$
The ADM Hamiltonian (2.18) now takes the form, using (2.29-31)
$$H=g^{1/2}={\a^2\over 4}\sin{2t\over \a}(r_1^-r_2^+ - r_1^+r_2^-) =
{\a \over {2\sqrt{\tau^2 - 4\Lambda}}}(r_1^-r_2^+ - r_1^+r_2^-)
\eqno(2.36)$$
and generates the $\tau$ development of the modulus (2.29) and momentum 
(2.30) through
$${d\pi\over d\tau}= \{\pi,H\},\qquad {dm\over d\tau}=\{m,H\}\eqno(2.37)
$$
Alternatively, the Hamiltonian
$$ H^\prime={d\tau\over dt}H={4\over {\a^2}}\csc^2 {2t\over \a}H 
$$
generates evolution in coordinate time $t$ by 

$${d\pi \over dt}=\{\pi, H^\prime\},\qquad {dm\over dt}=\{m, H^\prime\}  
$$

Using (2.29-32) one can show that in terms of the holonomy parameters 
$r_{1,2}^{\pm}$ the constants of the motion  $C_1  - C_6$ (2.22) 
are particularly simple 

$$ C_1 = \um ({(r_1^+)}^2 + {(r_1^-)}^2)$$
$$ C_2 = \um ({(r_2^+)}^2 + {(r_2^-)}^2)$$
$$ C_3 = \um (r_1^+r_2^+ + r_1^-r_2^-)$$
$$ C_4 = {\a \over 4} ({(r_1^-)}^2 - {(r_1^+)}^2)$$
$$ C_5 = {\a \over 4} ({(r_2^+)}^2 - {(r_2^-)}^2)$$
$$ C_6 = {\a  \over 4}(r_1^-r_2^- - r_1^+r_2^+) \eqno(2.38)$$
or alternatively, from (2.9)
$$C_1^{\pm} = (r_1^{\mp})^2$$
$$C_2^{\pm} = (r_2^{\mp})^2$$
$$C_3^{\pm} = r_1^{\mp}r_2^{\mp}\eqno(2.39)$$
and they are evidently time independent.
It can easily be checked from (2.38) that the constants  $C_1  
- C_6$ are not all independent. They satisfy 

$$C_{2}C_{4} - C_{1}C_{5} - C_{3}C_{6} = 0  \eqno(2.40)$$
and the time independent part $\bar H$ (2.19) of the ADM Hamiltonian 
(2.18), where
$$\bar H = {\a \over 2}(r_1^-r_2^+ - r_1^+r_2^-)\eqno(2.41) $$
 is expressed by  either
$$\Lambda {\bar H}^2 = (C_3)^2 - C_1 C_2 \quad or  \quad {\bar H}^2 
= (C_6)^2 + 4 C_4 C_5\eqno(2.42)$$
 
\vskip 0.7truecm

\centerline{\bf III. THE ANTI-DE SITTER ALGEBRA}
\vskip 0.7truecm

In [8] it was shown that suitable combinations of these six constants of 
the motion satisfy the Lie algebra of the Poincar\'e group%
\footnote{\raise2pt \hbox{\ninerm*}}{\ninerm The sign of $P_0$ reported 
in [8] was incorrect. The version here is the correct one.}. Here the 
combinations that satisfy the Lie algebra of the anti-de Sitter group 
$\hbox{SO}(2,2)$  are
$$P_0 = - \um (C_1 + C_2), P_1 = \um (C_1 - C_2), P_2 =C_3 $$
$$J_{12} =C_5 - C_4, J_{02} = C_4 + C_5, J_{01} = - C_6  \eqno(3.1)$$
 that is
$$\{J_{ab}, J_{cd}\} = \n_{ac} J_{bd} - \n_{bc} J_{ad} -\n_{ad}  
J_{bc}+
\n_{bd} J_{ac}$$
$$\{P_a , P_b\} = \La J_{ab}$$
$$\{J_{ab}, P_c\} = \n_{ac}P_b - \n_{bc}P_a  \eqno(3.2)$$
where 

$$a,b,c = 0,1,2, \quad \n_{ab} = diag (-1,1,1)$$
 satisfying
$$ P_a J_{bc} \ep^{abc} =0, \quad \ep^{012}= - \ep_{012}=1   
\eqno(3.3)$$
The classical algebra can be checked in either set of variables 
using the Poisson 
brackets (2.28) or (2.35) though it is evidently somewhat easier in the 
holonomy variables.

It is useful to define the generators
$$ {j_a}^{\pm}=  \um \ep_{abc}J^{bc} \pm \a P_a  \eqno(3.4)$$
with each  $\pm$ copy satisfying the Lie algebra of 
$\hbox{so}(1,2) \approx \hbox{sl(2,$\IR$)}$.
$$\{j_a^{\pm},j_b^{\pm}\} = 2\ep_{abc} {j^c}^{\pm}$$
$$\{{j_a}^+, {j_b}^-\} =0  \eqno(3.5)$$
with 
$$ j= {j_a}^+ {j^a}^+ = {j_a}^- {j^a}^- =0 
\eqno(3.6)$$
Explicitly we have
$$j_0^{\pm} = \mp {\a \over 2}(C_1^{\mp} + C_2^{\mp})= 
\mp {\a \over 2} ({(r_1^{\pm})}^2 + {(r_2^{\pm})}^2)$$
$$j_1^{\pm} =\mp {\a \over 2} (C_2^{\mp} - C_1^{\mp})= 
\pm {\a \over 2} ({(r_1^{\pm})}^2 - {(r_2^{\pm})}^2)$$
$$j_2^{\pm} = \pm \a C_3^{\mp} = \pm \a r_1^{\pm}r_2^{\pm}  \eqno(3.7)$$
Note that $j_a^+$ depends only on the $r^+$'s and $j_a^-$ 
only on the $r^-$'s.
The time independent part $\bar H$ (2.41) of the 
ADM Hamiltonian $H$ (2.18) or (2.36) is assumed to be positive. This 
guarantees that the imaginary part 
$m_2$ of the modulus (2.29) is also positive in the range 
$t~ \epsilon~ (0, {\pi \a \over 2})$, since, from (2.29) 
$$m_2 = e ^{-q^1}= \sin{2t\over \a}{(r_1^-r_2^+ - r_1^+r_2^-) 
\over {\vert r_2^-e^{it/\a} + r_2^+e^{-{it/\a}} \vert}^2 }.$$

$\bar H$ can also be expressed, from (2.42), (3.1) and (3.7), in terms 
of the anti-de Sitter and $\hbox{sl(2,$\IR$)}$
generators
$${\bar H}^2 = - \um J_{ab}J^{ab} = \um j_a^+ {j^a}^-\eqno(3.8) $$
$$\La {\bar H}^2= P_aP^a  \eqno(3.9)$$
\vskip 0.7truecm

\centerline{\bf IV.QUANTUM THEORY}
\vskip 0.7truecm
\vskip 0.7truecm

\centerline{\bf A. ADM Quantisation}
\vskip 0.7truecm

To quantise the ADM-reduced dynamics one can proceed as suggested in [1] 
and developed in detail in [2] for the vacuum case. Indeed the only 
essential difference between the reduced Schr\"odinger dynamics here and 
that for $\Lambda = 0$ is that the relationship between the time 
coordinate $t^{\prime} $ (c.f. (2.20)) and the mean curvature depends upon 
$\Lambda$ -the reduced Hamiltonian operator is independent of $\Lambda$.

A "choice" which must be made in either case is whether to formulate the 
reduced quantum mechanics on the full Teichm\"uller space for the torus 
(i.e. the 2-dimensional hyperbolic space with global coordinates $q^1, 
q^2 $ and Riemannian metric $(dq^1)^2 + e^{2q^1}(dq^2)^2 $ or instead 
upon the moduli space obtained from quantising the Teichm\"uller space 
by the action of the modular group discussed below in Section VI. 

Since the latter choice implements invariance of the physical states 
with respect to "large" diffeomorphisms as well as the small ones which 
are connected to the identity, it seems to be the natural one to make. 
The Schr\"odinger Hamiltonian determined from (2.19) may, as before, be 
defined as the positive square root of the invariant Laplacian defined 
on moduli space - a choice which seems to lead to well-defined quantum 
dynamics for the full physically desirable range of $t^{\prime} $ [2]. 

However, as we shall show below in a related context, the conserved 
quantities $C_1  - C_6$ are not invariant with respect to the 
transformations generating the modular group (c.f. (6.3)). Thus one does 
not expect to be able to implement them globally as self-adjoint 
operators in fully reduced moduli-space quantisation, a conclusion which 
seems to have been reached by a more geometrical line of reasoning by 
Hajicek as well [11].

One could perhaps try to implement the operator analogues of $C_1  - C_6$
on a partially reduced quantisation, working on Teichm\"uller space 
instead of moduli space, but unfortunately there is no known ordering of 
the operator analogues of expressions (2.22) which captures the 
$\hbox{so}(2,2)$ algebra expected from the classical considerations of 
Section III. Fortunately, however, this problem disappears when the 
conserved quantities are instead quantised in terms of the holonomy 
parameters, as we shall now show.

\vskip 0.7truecm
\centerline{\bf B. Holonomy parameter quantisation}
\vskip 0.7truecm

The quantisation of the algebra (3.2) is straightforward in terms 
of the  holonomy 
parameters. Indeed if all the $r_{1,2}^{\pm}$ are promoted to  
operators 
${\hat r}_{1,2}^{\pm}$ satisfying the commutators
$$[{\hat r}_1^{\pm}, {\hat r}_2^{\pm}] = 
{\hat r}_1^{\pm} {\hat r}_2^{\pm} -  {\hat r}_2^{\pm} {\hat r}_1^{\pm}
= \mp {i {\hbar} \over \a}$$
$$[{\hat r}^{\pm},{\hat r}^{\mp}] = 0\eqno(4.1)$$
then there are no ordering problems in $\hat H$ or 
$\hat {\bar H}$ (3.8), that is,
$$\hat H = {\a\over 2}\sin{2t\over \a}\hat {\bar H }, \quad 
\hat {\bar H} = {\a \over 2} (\hat r_1^-\hat r_2^+ 
- \hat r_1^+\hat r_2^-)\eqno(4.2)$$
 and the moduli and 
momenta, {\it ordered as in} (2.29-30) , that is,
$$\hat m = \left(\hat r_1^-e^{it/\a} + \hat r_1^+e^{-{it/\a}}\right) 
 \left(\hat r_2^-e^{it/\a} + \hat r_2^+e^{-{it/\a}}\right)^{\lower2pt%
 \hbox{$\scriptstyle -1$}}\eqno(4.2)$$
$$\hat \pi = -{i\a\over 2\sin{2t\over \a}}\left(\hat r_2^+e^{it/\a} 
 + \hat r_2^-e^{-{it/\a}}\right)^{\lower2pt\hbox{$\scriptstyle 
2$}}\eqno(4.4)$$
satisfy
$$[\hat m^\dagger,\hat \pi]=[\hat m,\hat \pi^\dagger]= - 2 i\hbar, \qquad
  [\hat m,\hat \pi] = [\hat m^\dagger,\hat \pi^\dagger] = 0\eqno(4.5)$$
$$[\hat \pi, \hat H] = i \hbar {d\hat \pi\over d\tau},
  \qquad  [\hat m, \hat H] = i \hbar {d\hat m\over d\tau}\eqno(4.6)$$
which follow from the commutators (4.1).

For the $\hbox{sl(2,$\IR$)}$ generators (3.7)
it is clear that there are no ordering problems in $j_0^{\pm}$ or 
$j_1^{\pm}$ that is
$${\hat j}_0^{\pm} = \mp {\a \over 2} ({({\hat r}_1^{\pm})}^2 
+ {({\hat r}_2^{\pm})}^2)$$
$${\hat j}_1^{\pm} = \pm {\a \over 2} ({({\hat r}_1^{\pm})}^2 
- {({\hat r}_2^{\pm})}^2)\eqno(4.7)$$

whereas in $j_2^{\pm}$ the symmetric ordering 

$${\hat j}_2^{\pm} = \pm {\a\over 2}
({\hat r}_1^{\pm}{\hat r}_2^{\pm} + {\hat r}_2^{\pm}{\hat r}_1^{\pm})   
\eqno(4.8)$$
will give the commutators
$$[{\hat j}_a^{\pm},{\hat j}_b^{\pm}] = 2i {\hbar}  
\epsilon_{abc}{{\hat j}^c}{}^{\pm}$$
$$[{{\hat j}_a}^+,{{\hat j}_b}^-] = 0  \eqno(4.9) $$
The commutator (4.1)  defines a spinor norm
$$\ep^{AB}{\hat r}_B^{\pm}{\hat r}_A^{\pm}={{\hat r}^A}{}^{\pm}
{\hat r}_A^{\pm}= [{\hat r}_2^{\pm}, {\hat r}_1^{\pm}] = 
\pm {i {\hbar} \over \a}  \eqno(4.10)$$
with $\ep ^{12} = - \ep ^{21} = 1$ and 
${\hat r}^A{}^{\pm}=\ep ^{AB}{\hat r}_B^{\pm}$. 

The quantum Casimir (3.6) commutes with all the ${\hat j}_a^{\pm}$, and
is no longer zero, but $O(\hbar^2)$ 

$${\hat j} ={{\hat j}_a}^{\pm} {{\hat j}^a}{}^{\pm}= {3{\hbar}^2 \over  
4}  \eqno(4.11)$$
Similarly the identity (3.8) acquires $O(\hbar^2)$ corrections
$${\hat {\bar H}}^2 = \um{{\hat j}_a}^+ {{\hat j}^a}{}^- 
+ {{\hbar}^2 \over 2}   
\eqno(4.12)$$
This particular value of the Casimir (4.11)
corresponds to a particular discrete representation of $SU(1,1)$ in 
which ${\hat j}$ and ${{\hat j}_0}^{\pm}$ are diagonal. This will be 
discussed elsewhere.
Note that the only ordering ambiguity is in ${{\hat j}_2}{}^{\pm} $ 
(4.8) but that any other ordering would only produce terms of $O(\hbar^2)$ 
on the R.H.S. of (4.9).
\vskip 0.7truecm

\centerline{\bf V. EXTENDED QUANTUM ALGEBRA}
\vskip 0.7truecm

The two quantum  $\hbox{so}(1,2)$  algebras introduced in the previous 
section can 
be extended to $\hbox{so}(2,3)$  on inclusion of the time independent, 
constant part $\hat {\bar H}$ (4.2) of the ADM Hamiltonian as follows.

The ADM Hamiltonian is not a constant of the motion, in fact classically 
it represents the surface area of the torus which increases from zero 
(the initial singularity) to a maximum and then recollapses. In this 
section we work only with the time independent part $\hat {\bar H}$ 
(4.2) since the time dependence is simply a multiplicative factor,
positive in the range $t\epsilon (0, {\pi \a \over 2} )$. 
The positivity,
classically, of the constant, global part is clearly  related 
to the question of 
ranges and signs of the classical, constant, holonomy  parameters 
${r_{1,2}}^{\pm}$. We assumed that
$r_1^-r_2^+ - r_1^+r_2^- > 0$ which guarantees that, classically, 
$m_2 = e ^{-q^1} >0$. As a quantum mechanical operator 
$ \hat {\bar H}$ we cannot
guarantee that its spectrum be positive definite 
without assuming some representation for the operators satisfying the 
commutators (4.1). This problem is outside the scope of this paper.

It can be 
checked from (4.2) and (4.7-8) that with 
$${\hat J}_{ab} = - \um \ep_{abc} ({{\hat j}^c}{}^+ + {{\hat j}^c}{}^-)  
\qquad {\hat P}_a =  {{{\hat j}_a}^+ - {{\hat j}_a}^- \over 
2\a}\eqno(5.1)$$
and using the commutators (4.1) it follows that
$[\hat {\bar H}, {\hat J}_{ab}]=0 $, whereas 
$[\hat {\bar H} , {\hat P}_a]\ne 0$ 
but instead defines a new constant three-vector ${\hat v}_a$ by
$$[\hat{\bar H} , {\hat P}_a] = i\hbar\a~ {\hat v}_a, a=0,1,2\eqno(5.2)$$
where
$${\hat v}_0=-{\a \over 2}({\hat r}_1^+ {\hat r}_1^- + {\hat r}_2^+ 
{\hat r}_2^-)$$
$${\hat v}_1={\a \over 2}({\hat r}_1^+ {\hat r}_1^- - {\hat r}_2^+ 
{\hat r}_2^-)$$
$${\hat v}_2=-{\a \over 2}({\hat r}_1^+ {\hat r}_2^- + 
{\hat r}_2^+ {\hat r}_1^-)\eqno(5.3)$$
The ${\hat v}_a, \hat {\bar H} $ {\it classically} form a null vector 
$${\hat v}_a{\hat v}^a ={\hat {\bar H}}^2 - {\hbar^2 \over 2}
\eqno(5.4) $$ 
as can be seen by expressing their components in terms of the {\it 
commuting} spinors
$${\hat r}^{\pm} = {{\hat r}_1^{\pm}\choose {\hat r}_2^{\pm}}\eqno(5.5)$$
$${\hat v}_0 = - {\a \over 2} {{\hat r}^+}{}^T \I {\hat r}^- $$
$${\hat v}_1 =  {\a \over 2} {{\hat r}^+}{}^T \sg_3 {\hat r}^-$$
$${\hat v}_2 =  {\a \over 2} {{\hat r}^+}{}^T \sg_1 {\hat r}^- $$
$$\hat {\bar H} = - i {\a \over 2} {{\hat r}^+}{}^T \sg_2 {\hat r}^- 
 \eqno(5.6)$$
where the $\sg_{1,2,3}$ are the usual Pauli matrices and, from (4.1)
$$  [{\hat r}^+,{\hat r}^-] =0  $$

Note that the above vector ${\hat v}_a$ and $\hat {\bar H}$ 
require {\it both } the ${\pm}$ spinors (5.5). This is in contrast 
to the generators  ${\hat j}_a^+$ and ${\hat j}_a^-$ (4.7-8)
of the two commuting $\hbox{sl(2,$\IR$)}$
subalgebras (4.9).

The extended algebra of the {\it ten} $\hat {\bar H}, {\hat j}_a^{\pm}, 
{\hat v}_a, a=0,1,2$ closes as follows
$$[{\hat j}_a^{\pm},{\hat j}_b^{\pm}] = 2i {\hbar}  
\epsilon_{abc}{{\hat j}^c}{}^{\pm}$$
$$[{{\hat j}_a}^+,{{\hat j}_b}^-] = 0  \eqno(5.7) $$

$$[\hat {\bar H},{\hat v}_a] = i\hbar \a {\hat P}_a = 
{i\hbar \over 2}({\hat j}_a^+ - {\hat j}_a^-)\eqno(5.8)$$
$$[\hat {\bar H},{\hat j}_a^{\pm}] = \pm i \hbar {\hat v}_a\eqno(5.9)$$
$$[{\hat v}_a,{\hat v}_b] = -{i \hbar \over 2} \epsilon_{abc}
({{\hat j}^c}{}^+ + {{\hat j}^c}{}^-)\eqno(5.10)$$
$$[{\hat j}_a^{\pm},{\hat v}_b] = \mp i \hbar \eta_{ab} \hat {\bar H} + 
i \hbar \epsilon_{abc} {\hat v}^c\eqno(5.11)$$
with the identities
$${\hat v}^a{\hat j}_a^{\pm} = {\hat j}_a^{\mp}{\hat v}^a = 
\pm {3i\hbar \over 2}\hat {\bar H}\eqno(5.12) $$
in addition to  (4.11),(4.12) and (5.4), making a total of 6 identities.

The above 10-dimensional algebra is isomorphic to the Lie algebra 
of $\hbox{so}(2,3)$, whose corresponding group is the conformal group of 
3-dimensional Minkowski space. The dilatation $D$ is to be 
identified with - $ \hat {\bar H}$, the translations with ${\hat P}_a^-$,
 and the conformal accelerations are denoted by ${\hat P}_a^+$, where
$${\hat P}_a^{\pm} = \a {\hat P}_a \pm {\hat v}_a $$
\vskip 0.7truecm

\centerline{\bf VI. THE QUANTUM MODULAR GROUP}
\vskip 0.7truecm

The modular group acts classically on the torus modulus and momentum as

$$\eqalign{&S: m\rightarrow 
  - m^{-1},\qquad p\rightarrow{\bar m}^2 p\cr
  &T: m\rightarrow m+1,\qquad p\rightarrow p\cr}\eqno(6.1)$$

and is equivalent to its action on the holonomy parameters
$$\eqalign{&S:r_1^{\pm}\rightarrow r_2^{\pm},\qquad 
    r_2^{\pm}\rightarrow - r_1^{\pm}\cr
&T:r_1^{\pm}\rightarrow r_1^{\pm} + r_2^{\pm},\qquad  
    r_2^{\pm}\rightarrow r_2^{\pm} ,\cr}\eqno(6.2)$$

Either (6.1) or (6.2) can be used to check that the Hamiltonian and 
Poisson brackets are invariant, and that the constants  $C_1  
- C_6$ transform as
$$\eqalign{S:&C_1\rightarrow C_2, \quad C_2\rightarrow C_1, \quad
               C_3\rightarrow - C_3,\cr
&C_4\rightarrow C_5, \quad
C_5\rightarrow C_4, \quad
C_6\rightarrow - C_6\cr
T:&C_1\rightarrow C_1 + C_2 + 2C_3, \quad
C_2\rightarrow C_2, \quad
C_3\rightarrow C_2 + C_3,\cr
&C_4\rightarrow C_4 - C_5 + C_6, \quad
C_5\rightarrow C_5, \quad
C_6\rightarrow C_6 - 2C_5\cr}  \eqno(6.3)$$
whereas  $v_a$ and the $\hbox{SO}(1,2)$  generators $j_a^{\pm}$ 
transform as
$$\eqalign{S:&j_0^{\pm}\rightarrow j_0^{\pm}\cr
&j_1^{\pm}\rightarrow - j_1^{\pm}\cr
&j_2^{\pm}\rightarrow - j_2^{\pm}\cr
T:&j_0^{\pm}\rightarrow {3 \over 2}j_0^{\pm} + \um j_1^{\pm} - 
j_2^{\pm}\cr
&j_1^{\pm}\rightarrow - \um j_0^{\pm} + \um j_1^{\pm} +
j_2^{\pm}\cr
&j_2^{\pm}\rightarrow - j_0^{\pm} - j_1^{\pm} + j_2^{\pm}\cr}   
\eqno(6.4)$$

With the ordering of (4.2) (the only ambiguity), the quantum action 
of the modular group is the
same as the classical one, with no $O(\hbar)$ corrections, and is
generated by the $\hbox{SO}(2,2)$  anti-de Sitter subgroup by 
conjugation with the operators $U_T $  and $U_S$ where

$$U_T= \exp {{i \over 2\hbar}(j_0^{\pm} +j_1^{\pm})}= 
\exp {\mp{i\a \over 2\hbar}C_2^{\mp}} 
= \exp {\mp {i\a \over 2\hbar}(r_2^{\pm})^2}\eqno(6.5)$$
$$U_S = \exp {{i\pi \over 2\hbar}j_0^{\pm}}= 
\exp {\mp{i\pi\a \over 4\hbar}(C_1^{\mp} + C_2^{\mp})} 
= \exp {\mp {i\pi\a \over 4\hbar}((r_1^{\pm})^2) 
+(r_2^{\pm})^2)}\eqno(6.6)$$

The first of these (6.5) appeared in [7] in a different notation. 
The second (6.6) was calculated 
independently by one of us (J.E.N.) and S. J. Carlip. 
 The remaining 
generator $j_2^{\pm} = \pm C_3^{\mp}$ acts, with parameter $\epsilon$, 
on the holonomy parameters as
$$\exp ({{i \ep \over \hbar}j_2^{\pm}})~ r_1^{\pm}~\exp ({- 
{i \ep \over \hbar}j_2^{\pm}}) =  r_1^{\pm}~\exp {- \ep}~$$
$$\exp ({{i \ep \over \hbar}j_2^{\pm}})~ r_2^{\pm}~\exp ({- 
{i \ep \over \hbar}j_2^{\pm}}) =  r_2^{\pm}~\exp {\ep}\eqno(6.7)
$$
so that the moduli and their momenta scale as 
$$ m \to m~\exp {- 2\ep}, p \to p~\exp {2\ep} \eqno(6.8)$$

It can be checked that, using (6.2), the commutators (4.1) and therefore
the quantum algebra of Section V, and the identities (4.11-12), (5.4) 
and (5.12) are invariant.
\vskip 0.7truecm

\centerline {\bf VII. CONCLUSION}
\vskip 0.7truecm

We have described several different (and possibly inequivalent) 
quantisations of 2+1 - dimensional gravity on $\IR \times T^2$ in the 
presence of a negative cosmological constant. The direct ADM approach 
leads naturally to a well-defined Schr\"odinger dynamics (with positive 
definite, self-adjoint Hamiltonian operator) but does not succeed in 
exploiting, or even implementing, operator analogues of the classically 
independent conserved quantities arising from traces of holonomies. 
Perhaps this is not surprising since such a result (effectively a 
solution of the Heisenberg equations of motion) would encode within it 
the details of the spectrum of the Laplace-Beltrami operator on moduli 
space- a heretofore unsolved problem. For this same reason we suspect 
that the elegant explicit solution of a different formulation of the 
Heisenberg equations of motion given in [5] is not unitarily equivalent 
to the solution of the Schr\"odinger equation involving the square root 
of the Laplace-Beltrami operator (which was studied in detail in [2]). 
Indeed a representation for the fundamental quantised holonomy 
parameters which guarantees positivity of the Carlip-Nelson form (2.36 
and 4.2) of the reduced Hamiltonian (2.18) does not seem to be known. 

On the other hand if we set aside the above questions and quantise 
directly in terms of the holonomy parameters we can easily order the 
generators of the $\hbox{so}(2,2)$ algebra, or, more generally, those 
of the $\hbox{so}(2,3)$ algebra, so 
as to implement these algebras quantum mechanically. Within this same 
context we also formulate the action of the quantum modular group as 
generated by a certain discrete subgroup of the $\hbox{SO}(2,2)$ 
anti-de Sitter group.

The conserved quantities  $C_1  
- C_6$ are shown explicitly to be global
and time independent by expressing them in terms of the parameters of 
the traces of holonomies through a time dependent canonical 
transformation. Certain combinations of them satisfy the Lie algebra of 
the anti- de Sitter group $\hbox{SO}(2,2)$. 
Quantisation is straightforward in terms of the holonomy 
parameters. When the Hamiltonian is 
included three new (quantum) conserved quantities are found and the 
algebra extends to that of the conformal group $\hbox{SO}(2,3)$.
The quantum modular group is generated by the anti- de Sitter subgroup, 
namely by these (quantised) conserved quantities. The quantum 
modular group appears as a discrete subgroup of the conformal group.

The group extension, that is, the Hamiltonian and the vectors $v_a$ act 
differently by mixing the two $\hbox{so}(1,2)$ algebras.
The role of these 
operators and their action on the quantum states of the system is under 
investigation.

A related construction for zero cosmological constant using ADM 
variables can be found in 
[11]. There the constants found by one of us [8] are used as generators 
of isometries in the unreduced, ADM, Hamiltonian formalism.

For completeness we note that although we have only discussed the case 
of negative cosmological constant there would seem to be no obstruction 
to the discussion for $\La$ positive or zero (see the discussion in 
[5]). For example, for $\La > 0$, the parameters ${r_{1,2}}^{\pm}$ 
would be unchanged but the holonomies (2.24), the "shifted connections" 
(2.6), and in consequence, the corresponding $\hbox{sl(2,$\IC$)}$ 
generators ${j_a}^{\pm}, a=0,1,2$ would be complex conjugates of 
each other rather than real and independent as here, for $\La < 0$.

\vskip 0.7truecm

\centerline {\bf ACKNOWLEDGEMENTS}
\vskip 0.7truecm

This work was supported in part  
by INFN Iniziativa Specifica FI41, the European Commission TMR 
programme ERBFMRX-CT96-0045, the European Commission HCM programme 
CHRX-CT93-0362, the Erwin Schrodinger Institute (Vienna), and NSF 
grant PHY - 9503133.

\vskip 0.7truecm

\centerline {\bf REFERENCES}
\vskip 0.7truecm

\item{[1]\ } V. Moncrief, J. Math. Phys. {\bf 30}(12), 2907 (1989).
\item{[2]\ }R.~Puzio,  Class. Qu. Grav. {\bf 11}, 609-620 (1994).
\item{[3]\ }J.~E.~Nelson and T.~Regge, Nucl.\ Phys.\ {\bf B328}, 190 
(1989);J.~E.~Nelson,  T.~Regge and F.~Zertuche, Nucl.\ Phys.\ 
{\bf B339}, 516 (1990). 
\item{[4]\ }S.~Carlip, Phys.\ Rev.\ {\bf D42}, 2647 (1990); Phys.\ Rev.\ 
 {\bf D45}, 3584 (1992); Phys.\ Rev.\ {\bf D47}, 4520 (1993).
\item{[5]\ }S.~Carlip and J.~E.~Nelson, Phys. Lett. {\bf B 324}, 299 
(1994); Phys. Rev. {\bf D 51} 10, 5643 (1995).
\item{[6]\ }J.~E.~Nelson and T.~Regge, Commun.\ Math.\ Phys.\ 
{\bf 141}, 211 (1991); Commun.\ Math.\ Phys.\ {\bf 155}, 561 (1993).
\item{[7]\ }J.~E.~Nelson and T.~Regge, Phys.\ Lett.\ {\bf B272}, 213 (1991).
\item{[8]\ } V. Moncrief, J. Math. Phys. {\bf 31}(12), 2978 (1990);
\item{[9]\ }S.~Martin, Nucl.\ Phys.\ {\bf B327}, 178 (1989).
\item{[10]\ }K.~Ezawa, {\it Reduced Phase Space of the First Order 
Einstein Gravity on $\IR \times T^2$}, Osaka preprint No. OU-HET-185, 
1993.
\item{[11]\ }P. Hajicek, "Group-Theoretical Quantisation of 2+1 Gravity 
in the Metric-Torus Sector  " gr-qc/9703030.

\vfill\eject
\bye